\documentstyle[emulateapj]{article}
\singlespace
\journalid{}{}
\articleid{}{}

\slugcomment{accepted for publication in the {\it Astrophysical Journal Letters}}

\newcommand{\etal}{{\it et al.~}}
\newcommand{\fb}{{$f_B$}}
\newcommand{\g}{{\it g}}
\newcommand{\r}{{\it r}}
\newcommand{\igt}{{\it i}}
\newcommand{\lx}{{$L_X$}}
\newcommand{\blue}{{\it blue~}}

\newcommand{\CM}{{color-magnitude~}}
\newcommand{\HST}{{\it Hubble Space Telescope~}}
\newcommand{\BOe}{BO effect~}
\newcommand{\BOa}{Butcher \& Oemler~}

\begin{document}

\title{The Butcher-Oemler Effect in 295 Clusters: Strong 
Redshift Evolution and Cluster Richness Dependence}
\lefthead{Margoniner \etal}
\righthead{The Butcher-Oemler Effect in 295 Clusters}

\author{V. E. Margoniner \altaffilmark {1,2}}

\affil{vem@physics.bell-labs.com}

\author{R. R. de Carvalho \altaffilmark {2}}


\author{R. R. Gal \altaffilmark {3} and S. G. Djorgovski \altaffilmark {3}}


\altaffiltext{1}{Bell Laboratories, Lucent Technologies, Murray Hill, NJ 07974}
\altaffiltext{2}{Observat\'orio Nacional, CEP 20921-400, Rio de Janeiro, Brazil}
\altaffiltext{3}{Palomar Observatory, Caltech, MC105-24, Pasadena, CA 91125}

\begin{abstract}
We examine the Butcher-Oemler effect and its cluster richness
dependence in the largest sample studied to date: 295 Abell clusters.
We find a strong correlation between cluster richness and the fraction
of blue galaxies, \fb, at every redshift.  The slope of the $f_B(z)$
relation is similar for all richnesses, but at a given redshift, \fb\
is systematically higher for poor clusters.  This is the chief cause
of scatter in the \fb\ vs. $z$ diagram: the spread caused by the
richness dependence is comparable to the trend in \fb\ over a typical
redshift baseline, so that conclusions drawn from smaller samples have
varied widely.  The two parameters, $z$, and a consistently defined
projected galaxy number density, $N$, together account for all of the
observed variation in \fb\ within the measurement errors. The redshift
evolution of \fb\ is real, and occurs at approximately the same rate
for clusters of all richness classes.
\end{abstract}


\section{Introduction}

The Butcher-Oemler (BO) effect provided some of the first evidence for
the evolution of galaxies and clusters.  \BOa (1978, 1984) found an excess
of blue galaxies in high redshift clusters in comparison with the
typical early-type population observed in the central region of local
clusters (Dressler 1980). The \BOe has also been observed in more
recent studies, which indicate an even stronger evolution of the
fraction of blue galaxies in clusters (Rakos \& Schombert 1995,
Margoniner \& de Carvalho 2000).

The BO effect is an indicator of evolution in the cluster galaxy
population, which may result from changes in the morphology and
star-formation rates of member galaxies with redshift.  The fact that
blue galaxies are more commonly found at higher redshifts, and the
observation of an apparent excess of S0 galaxies in low redshift
clusters, lead to the suggestion by Larson \etal (1980) of an
evolutionary connection between S0 and spiral galaxies.  This idea can
explain the population of blue galaxies observed by Butcher \& Oemler
as spiral galaxies seen just before running out of gas, and the
disappearance of this population in more evolved, low redshift rich
clusters (Dressler \etal 1997, Couch \etal 1998).  In the last decade,
high-resolution \HST images have allowed the determination of the
morphology of these high redshift blue galaxies.  Dressler \etal
(1994), Couch \etal (1994, 1998), and Oemler \etal (1997) found that
the \BOa galaxies are predominantly normal late-type spirals, and that
dynamical interactions and mergers between galaxies may be a secondary
process responsible for the enhanced star formation.  A recent study
by Fasano \etal (2000) indicates that as the redshift decreases, the
S0 cluster population increases while the number of spiral galaxies
decreases, supporting Larson's original idea of spirals evolving into
S0s.

A second factor affecting the number of blue galaxies may be their
environment. The fraction of blue galaxies seems to be correlated with
local galaxy density, and with the degree of substructure in the
cluster.  Many studies have shown that the blue galaxies lie
preferentially in the outer, lower density cluster regions (Butcher \&
Oemler 1984, Rakos \etal 1997, Margoniner \& de Carvalho 2000).  Also,
Smail \etal (1998) studied 10 massive clusters and found that the
fraction of blue galaxies in these clusters is smaller than observed
in \BOa (1984) clusters at the same redshift range.  However, an
opposite trend is found in non-relaxed clusters with a high degree of
substructure, where the fraction of blue galaxies is higher than that
observed in regular clusters (Caldwell \& Rose 1997, Metevier \etal
2000).

The evolution of these galaxies is probably correlated with its
environment in the sense that spirals might consume and/or lose their
gas, evolving to SOs, while falling into the higher density cluster
regions.  Although the exact mechanism responsible for the \BOe is not
completely understood (Kodama \& Bower 2000), most authors seem to
agree that the effect is real.  A different idea is presented by
Andreon \& Ettori (1999) who show that the mean X-ray luminosity and
richness of \BOa (1984) clusters increases with redshift, and argue
that the increase in the fraction of blue galaxies with redshift may
not represent the evolution of a single class of objects.

We present observations of the \BOe in a large sample of 295 Abell
(1958, 1989) clusters of all richnesses, with no further selection on
the basis of morphology or X-ray luminosities. This is important
because all previous studies were based on samples biased toward
richer clusters, with some samples being further selected according to
morphology and/or X-ray luminosities, so that a combination of
different selection effects could be mimicking the observed $f_B(z)$
relation.  Although our sample inherits biases existent in the Abell
catalog, it should contain a more representative variety of clusters
(in terms of degree of substructure, richness, and mass) at each
redshift than any previous sample, and because of its large size
should allow a better determination of the relation.  Any results
driven by selection effects present in previous samples might become
apparent when compared with this one.  We describe the data in Section
2, the \BOe analysis in Section 3, and our conclusions in Section 4.

\section{Sample Selection and Input Data}

The data presented in this paper were obtained to calibrate the
DPOSS-II (the Digitized Second Palomar Observatory Sky Survey,
Djorgovski \etal 1999). It comprises 44 Abell clusters imaged at the
0.9m telescope at the Cerro Tololo Interamerican Observatory (CTIO)
(Margoniner \& de Carvalho 2000, hereafter MdC00), and 431 clusters
observed at the Palomar Observatory 1.5m telescope (Gal \etal 2000,
hereafter G00, and in preparation). The CCD images were taken in the
\g, \r~ and \igt~ filters of the Thuan \& Gunn (1976) photometric
system, with typical $1$-$\sigma$ magnitude errors of 0.12 in \g, 0.10
in \r, and 0.16 in \igt~ at $r=20.0^m$.  We have also observed 22
control fields in order to assess the background contribution. More
details concerning the data reduction, photometry, and catalog
construction can be found in MdC00 and G00.

From this original sample we excluded 120 clusters observed with a
small field of view CCD, and 31 very low redshift ($z<0.05$) clusters,
for which only the core region can be observed with our $13^{\prime}
\times 13^{\prime}$ images.  Also, 26 cluster fields which exhibited
galaxy counts comparable to the mean background, and one cluster with
a bright star occupying $\sim 25\%$ of the CCD region, were excluded
from the analysis.  Since 2 Abell clusters had repeated observations
from G00 and MdC00, our final sample comprises 295 clusters.  It is
important to note that at variance with previous studies, the sample
presented in this work is representative of all richness class
clusters (21\% $R=0$, 50\% $R=1$, 21\% $R=2$, and 8\% $R\ge3$). No
further selection regarding richness, morphology or degree of
sub-clustering was applied when determining our sample.  Spectroscopic
redshifts for 77 clusters were obtained from the literature, and for
the remainder photometric redshifts were estimated with an rms
accuracy of approximately 0.04 using the methodology described in
MdC00.

\section{Analysis of the Butcher-Oemler Effect}

The BO effect can be measured by comparing the fraction of blue
galaxies in clusters at different redshifts.  The $(g-r)~vs.~r$ \CM
(CM) relation was determined by fitting a linear relation to the red
galaxy envelope using the same prescription as in MdC00. Red envelopes
were subjectively classified as well-defined or not by visual
inspection.  Galaxies were defined as \blue if they had $(g-r)$ colors
$0.2^m$ below the linear locus in the CM relation.  We used the
spectral energy distribution of a typical elliptical galaxy (Coleman
\etal 1980) to derive $k(z)$ corrections for our sample, and also
corrected the data for extinction using the maps of Schlegel \etal
(1998).  Because the main purpose of this work is to study the
evolution of galaxy populations, fainter galaxies play a crucial role.
Whereas \BOa were limited to brighter galaxies by their photographic
data, our CCD sample allows us to probe significantly deeper,
selecting galaxies with \r~ magnitude between $M^*-1$ and $M^*+2$,
inside a region of radius $0.7$ Mpc around the cluster.  This fixed
linear size was chosen because our CCD images cover a field of radius
$\sim 0.5-4.0$ Mpc at $z=0.03-0.38$, and we want to study the same
physical region and the same galaxy luminosity range for the entire
sample.  We assume a cosmology with $H_{\circ} = 67$ Km s$^{-1}$
Mpc$^{-1}$, and $q_{\circ}=0.1$, in which $M_r^*=-20.16$ (Lin \etal
1996).

Most ($61\%$) of the clusters in our sample are at $z=0.1-0.2$, so
that the $0.7$ Mpc central region, and the entire luminosity range
between $M^*-1$ and $M^*+2$ can be observed. Clusters at higher
redshifts are limited at brighter absolute magnitudes and a correction
was applied in order to compare the fraction of blue galaxies in these
clusters with the rest of our sample.  The lower redshift clusters
suffer from the opposite problem, since the brighter galaxies will be
saturated, and we are also limited by the field of view. Details about
these corrections can be found in MdC00, the only difference being
that in the present work we adopt a more conservative brighter
limiting magnitude of $M^*+2$ instead of $M^*+3$.

The blue and total counts contain a mix of cluster members and
background galaxies.  Background variations place a fundamental limit
on the accuracy of \fb\ measurements: no matter how well one
determines the {\it mean} background from a number of control fields,
the background estimate for any particular cluster is no more accurate
than the {\it scatter} in the control fields.  For each cluster, we
measured the effect of this scatter by computing \fb\ using background
corrections from the 22 individual control fields scaled to the area
of the CCD and the CM relation appropriate to that cluster.  The final
\fb\ is given by the median and the rms is based on the central two
quartiles of the distribution.  Simple propagation of errors through
the equation $f_B \equiv {{n_{blue,cluster} -
n_{blue,background}}\over{n_{total,cluster} - n_{total,background}}}$
would give an overestimate because $n_{blue,background}$ and
$n_{total,background}$ are correlated.  The usual assumption that the
error in \fb\ is mainly due to Poisson statistics in the
background-corrected cluster counts does not properly take into
account the background variations and errors thus derived are about
three times smaller than ours when applied to the same data.

The final fractions of blue galaxies for the entire sample are shown
in the upper panel of Figure 1. For those clusters with multiple
observations, we have used the observation with smaller
$\sigma_{f_B}$.  Clusters with spectroscopic and photometric redshift
measurements are indicated by solid and open circles respectively.
The individual error bars are not presented to avoid confusion in the
plot, but the median $\sigma_{f_B}$ is 0.071.  The lower panel
presents only clusters that have: (1) $\sigma_{f_B}<0.071$, (2)
spectroscopic redshift measurements, and (3) a well-defined CM
relation.  The $1$-$\sigma_{f_B}$ errors are indicated for these
clusters.  In both panels, the solid lines indicates a linear fit
derived from the clusters with $z\leq0.25$ ($f_B=(1.24 \pm 0.07)z -
0.01$, with an rms scatter of $0.1$ for the entire sample shown in the
upper panel, and $f_B=(1.34 \pm 0.11)z - 0.03$, with an rms of $0.07$
for the sample shown in the lower panel). The fits were done with the
GaussFit program (Jefferys \etal 1988) taking into account the
measurement errors in \fb.  The dashed line indicates the rms scatter
around the fit, and while the upper panel shows a larger scatter, the
two derived relations are the same within the errors.  A clear trend
of strong evolution with redshift is seen.  A $\chi_{\nu}$ test
applied to the fit presented in the upper panel results in 1.36 if
only \fb~ errors are considered, and 1.13 when the uncertainties in
the photometric redshifts are also taken into account. Such small
$\chi_{\nu}$ is due however to large measurement errors in this
sample. A higher $\chi_{\nu}$ of 1.72 is obtained for the subsample of
clusters with smaller $\sigma_{f_B}$ and spectroscopic redshift (lower
panel of Figure 1).

Richness is a natural second parameter which might cause the range in
\fb\ at given redshift. The left upper panel of Figure 2 shows the
$f_B(z)$ diagram with symbol sizes scaled by $N$, the number of
galaxies between $M^*-1$ and $M^*+2$, inside $0.7$ Mpc, after
background correction.  Only clusters from the lower panel in Figure 1
and at redshifts between 0.1 and 0.2, where no corrections needed to
be applied to compute \fb, are presented in this figure.  The solid
line is the best-fit linear $f_B(z)$ relation for the 26 clusters
shown, and the dashed lines indicate the relations obtained when the
sample is subdivided in two 13 cluster samples according to
richness. The slopes of the three relations are the same within the
errors ($0.86\pm0.23$ for the entire sample, $0.96\pm0.26$ for the
subsample of richest clusters, and $0.86\pm0.56$ for the poorest
ones). Although the uncertainties are large, the rate of evolution is
approximately constant for all richnesses, and there is clearly an
effect, with richer clusters tending to lower \fb.  The suggestion by
Andreon \& Ettori (1999) that the observed increase of \fb~ with
redshift is caused by missing poor clusters at higher redshift is
therefore no longer tenable, because any such selection effects would
serve to {\it decrease} the redshift evolution of \fb.  The evolution
in \fb~ is real, and takes place in all richness classes.

To ensure that richness and redshift are not correlated in this
subsample, we plot $N~vs.~z$ in the right upper panel of Figure 2. The
best-linear fit ($N = -(38.2\pm106.2)z + (89.5\pm14.5)$) is indicated
by the solid line, and the dashed lines represent adding/subtracting
$1$-$\sigma$ uncertainties to its coefficients. Richness and redshift
are uncorrelated in this sample.

To gauge the richness importance, we investigate $f_B(z,N)$ relations
with various richness dependences ($N^{-1}$, $N^{-3/2}$, $N^{-2}$, and
$N^{-3}$), and found that $N^{-2}$ correlates slightly better with \fb.
The final best-fit relation for the data, when both redshift and
richness are considered, and taking into account the errors in \fb\
and $N$, is $f_B = (1.03\pm0.25)z + (388.3\pm111.4)N^{-2} - 0.04$,
with $\chi_{\nu}=0.99$.  The $F$ test indicates with $>99\%$ confidence
($F_{\chi_{\nu}}=23.8$, for 25 to 24 degrees of freedom) that richness
is responsible for the improvement observed in $\chi_{\nu}$. Richness
is therefore extremely important in determining \fb\ and it is in fact
enough to account for the scatter observed in a simple $f_B(z)$
relation.  In the lower left panel of Figure 2 we plot \fb\ {\it vs.}
$z + 378.6 N^{-2}$, which represent an edge-on view of the best-fit
plane $f_B(z,N)$, and in the lower right panel we present the redshift
dependence of a richness-corrected-\fb: the redshift evolution is
clear.
 
Although all 26 clusters shown in Figure 2 were used to compute the
relations indicated by solid lines in the figure, the two clusters
represented by open circles (Abell 520, and Abell 1081) seem to
deviate from the trend. These clusters also have the largest
$\sigma_N$ (derived from the scatter in the 22 control regions), and
if excluded from the fitting, the relation changes to $f_B = 1.13 z +
480.56 N^{-2} - 0.06$ ($\chi_{\nu}=0.95$), indicating a slightly
stronger redshift dependence.

Finally, we used 11 clusters with ROSAT X-ray luminosities to check
for correlations with \fb, but this small sample did not allow any
conclusions. There is a possible trend of increasing \lx\ with
redshift in the sense found by Andreon \& Ettori (1999) in BO's
original sample.  However, there is no clear trend of \fb\ with \lx,
arguing against any contamination of the $f_B(z)$ relation by
selection effects.  A comparison of the X-ray luminosities of these
clusters with their \fb\ could provide new insight to the BO effect,
and we are in the process of obtaining X-ray fluxes and upper limits
from the RASS (Rosat All Sky Survey) for all of these clusters, as
well as a larger sample of new clusters generated from DPOSS-II.

We stress that these results are not to be directly compared
with fractions of blue galaxies as originally defined in BO84.  The
most important reason for differences is the magnitude range used to
calculate \fb. The number of blue galaxies increases at fainter
magnitudes and it is therefore natural that our \fb\ measurements are
in general higher. Also, the usage of a rigid physical scale for all
clusters, instead of regions determined individually for each cluster
according to its density profile, should yield different \fb\
estimates.  When \fb\ is computed using the same absolute magnitude
range used by BO84, we find signs of evolution that are stronger than
originally suggested in their work, and which are consistent with
recent works by Rakos \& Schombert (1995) and MdC00. Limiting our
sample at brighter absolute magnitudes however increases the
$\sigma_{f_B}$ (and $\sigma_{N}$) because of the smaller galaxy count
statistics, and probes a smaller fraction of the cluster population.

\section{Summary and Conclusions}

We compute \fb\ for 295 randomly selected Abell (1989) clusters,
including galaxies as faint as $M^* + 2$ to sample a larger range of
the luminosity function and provide better statistics in each cluster.
The resulting \fb\ shows a stronger trend with redshift than did \fb\
as originally defined by BO84, consistent with the idea that the
\BOe is stronger among the late-type spirals and irregulars which
dominate the galaxy populations at intermediate and lower
luminosities.

This is the first sample large enough to allow the study of the \fb\
variations at a given redshift. A $\chi_{\nu}$ test applied to a
simple linear $f_B(z)$ relation shows that the scatter in \fb\ at
a given redshift is consistently larger than the measurement error,
indicating a real cluster-to-cluster variation.  We investigate the
richness dependence of the Butcher-Oemler effect, and find a strong
correlation between \fb\ and galaxy counts in the sense that poorer
clusters tend to have larger \fb\ than richer clusters at the same
redshift.  The inclusion of poor clusters tends therefore to increase
the slope of the $f_B(z)$ relation, and is another factor (together
with the inclusion of fainter galaxies) responsible for the stronger
evolution observed in this sample when compared to previous works
based mostly on rich clusters.

We show that the fraction of blue galaxies in a cluster can be
completely determined by its redshift and richness (within the
measurement errors). The evolution in \fb\ with redshift is real, and
occurs at approximately the same rate for clusters of all richnesses.

\acknowledgments

We thank D. Wittman, J.A. Tyson, A. Dressler, S. Andreon, and the
anonymous referee for very helpful comments and suggestions which
helped to improve the paper.  We also thank the Palomar TAC and
Directors for generous time allocations for the DPOSS calibration
effort, and numerous past and present Caltech undergraduates who
assisted in the taking of the data utilized in this paper.  RRG was
supported in part by an NSF Fellowship and NASA GSRP NGT5-50215.  The
DPOSS cataloging and calibration effort was supported by a grant from
the Norris Foundation.

\clearpage

\begin{figure}

\plotone{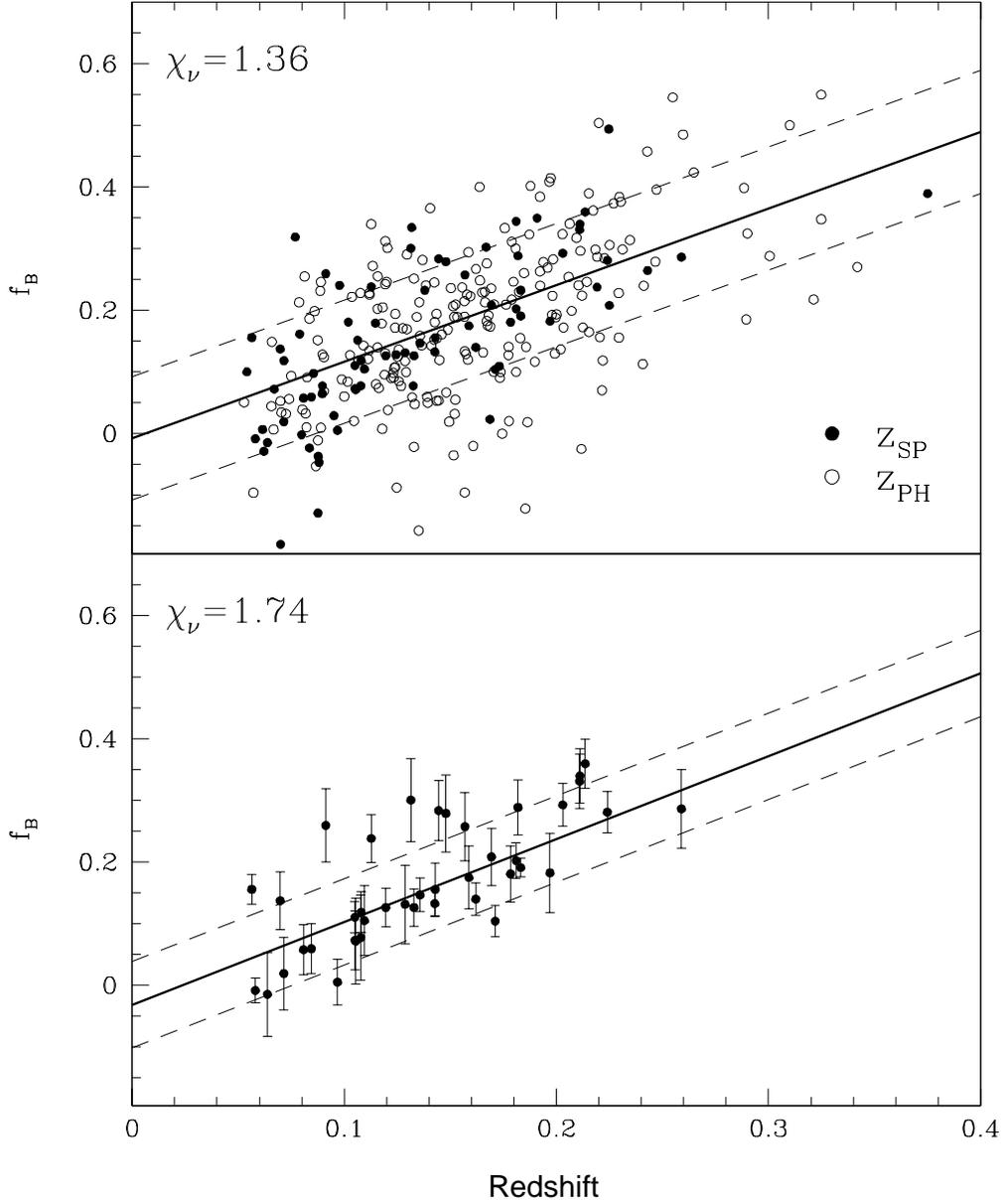}

\caption {Fraction of blue galaxies in the magnitude
range between $M^*-1$ and $M^*+2$, and within $0.7$ Mpc from the
center of the cluster.  The entire 295-cluster sample is shown in the
upper panel, and only clusters with (1) $\sigma_{f_B}<0.071$, (2)
spectroscopic redshift measurements, and (3) well-defined CM relation
are presented in the lower panel.  In each panel the solid line
indicates a linear $f_B(z)$ fit derived from the clusters with
$z\leq0.25$, and the dashed line represents the rms scatter of the
clusters around the fit.} 

\end{figure}

\clearpage

\begin{figure}
\plotone{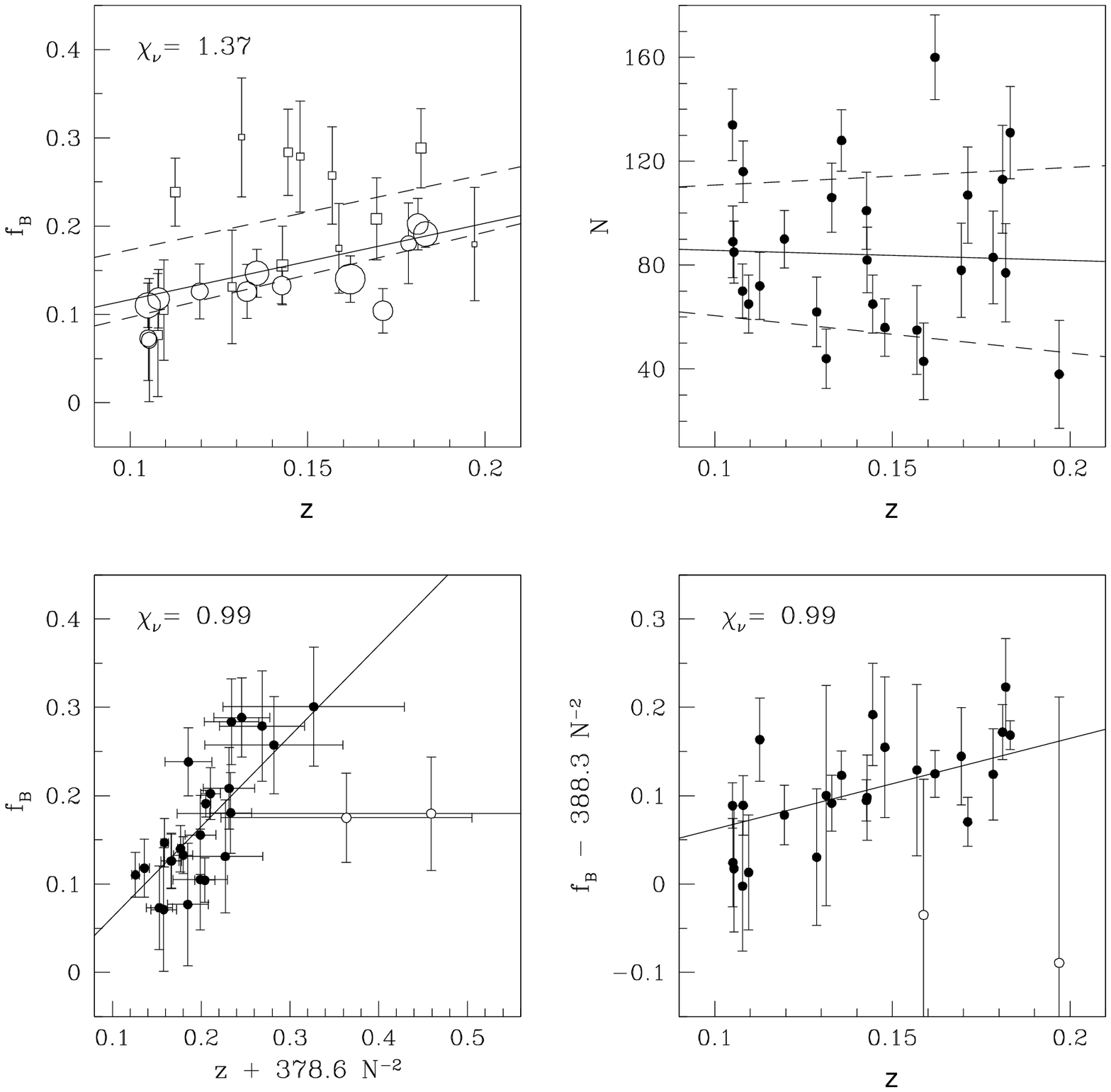}

\caption {Subsample of clusters with $0.1 \le z \le
0.2$ from the lower panel in Figure 1. The solid line in each panel
represents the best-fit to the entire sample. Left upper panel:
$f_B(z)$ diagram with marker sizes scaled by $N$ (galaxy counts). The
dashed lines indicate the fits computed separately for the richest
(circles) and poorest (squares) clusters. Right upper panel:
$N~vs.~z$. The linear $N(z)$ fit is indicated by the solid line, and
the dashed lines represent adding/subtracting $1$-$\sigma$
uncertainties to its coefficients.  Lower left panel: \fb\ {\it vs.}
$z + 378.6 N^{-2}$ (edge-on view of the best-fit plane $f_B =
(1.03\pm0.25)z + (388.3\pm111.4)N^{-2} - 0.04$ to the data). Lower
right panel: richness-corrected \fb\ as a function of
redshift.}

\end{figure}


\begin{thebibliography}{}

\bibitem[]{} Abell G.O., 1958, \apjs, 3, 211.
\bibitem[]{} Abell G.O., Corwin H.G., Olowin R.P., 1989, \apjs, 70, 1.
\bibitem[]{} Andreon S., Ettori S., 1999, \apj, 516, 647.
\bibitem[]{} Butcher H. \& Oemler A.Jr., 1978, \apj, 219, 18.
\bibitem[]{} Butcher H. \& Oemler A.Jr., 1984, \apj, 285, 426.
\bibitem[]{} Caldwell N. \& Rose J.A., 1997, \aj, 113, 492.
\bibitem[]{} Coleman G.D., Wu C-C., Weedman D.W. 1980, \apjs, 43, 393.
\bibitem[]{} Connolly A.J., Csabi I., Szalay A.S., 1995, \aj, 110, 6.
\bibitem[]{} Couch W.J., Ellis R.S., Sharples R.M., Smail I., 1994,
\apj, 430, 121.
\bibitem[]{} Couch W.J., Barger A.J., Smail I., Ellis R.S., Sharples
R.M., 1998, \apj, 497, 188.
\bibitem[]{} Djorgovski S.G., Gal R.R., Odewahn S.C., de Carvalho
R.R., Brunner R., Longo G., and Scaramella R., 1999, in Wide Field
Surveys in Cosmology, eds. S. Colombi, Y. Mellier, and B. Raban, Gif
sur Yvette: Eds. Fronti\`eres p. 89.
\bibitem[]{} Dressler A., 1980, \apj, 236, 351.
\bibitem[]{} Dressler A., Oemler A.Jr., Butcher H.R., Gunn J.E., 1994,
\apj, 430, 107.
\bibitem[]{} Dressler A., Oemler A.Jr., Couch W.J., Smail I., Ellis
R.S., Barger A., Butcher H.R., Poggianti B.M., Sharples R.M, 1997,
\apj, 490, 577.
\bibitem[]{} Fasano G., Poggianti B.M., Couch W.J., Bettoni D., Kj\ae
rgaard P., Moles M., 2000, \apj, in press.
\bibitem[]{} Gal R.R., de Carvalho R.R., Brunner R., Odewahn S.C.,
Djorgovski S.G., 2000, \aj, 120, 540.
\bibitem[]{} Jefferys, W.H., Fitzpatric M.J., McArthur B.E., 1988
Celest. Mech., 41, 39
\bibitem[]{} Kodama T. \& Bower R.G., 2000, submitted to \mnras.
\bibitem[]{} Larson R.B., Tinsley B.M., Caldwell N., 1980, \apj, 237,
692.
\bibitem[]{} Lin H., Kirshner R.P., Shectman S.A., Landy S.D., Oemler
A., Tucker D.L., Shechter P.L., 1996, \apj, 464, 60.
\bibitem[]{} Margoniner V.E. \& de Carvalho R.R., 2000, \aj, 119,
1562.
\bibitem[]{} Metevier A.J., Romer A.K., Ulmer M.P. , 2000 \aj, 119,
1090.
\bibitem[]{} Oemler A.Jr., Dressler A., Butcher H.R., 1997, \apj, 474,
561.
\bibitem[]{} Rakos K.D., Schombert J.M., 1995, \apj, 439, 47.
\bibitem[]{} Rakos K.D., Odell A.P., Schombert J.M., 1997, \apj, 490,
201
\bibitem[]{} Schlegel D.J., Finkbeiner D.P., Davis M., 1998, \apj,
500, 525.
\bibitem[]{} Smail I., Edge A., Ellis R., Blandford R., 1998, \mnras,
293, 124.
\bibitem[]{} Thuan T.X., Gunn J.E., 1976, \pasp, 88, 543.
\end{thebibliography}
\end{document}